\documentclass{PoS}
\usepackage{amsmath}
\usepackage{subfigure}
\DeclareMathOperator{\Tr}{{\rm Tr}}

\renewcommand{\Re}{\operatorname{Re}}

\title{Effective lattice theory for finite temperature Yang-Mills}

\ShortTitle{Effective lattice Polyakov loop theory for finite temperature Yang-Mills}

\author{\speaker{G.~Bergner}, O.~Philipsen\\
Institut f\"ur Theoretische Physik, Johann Wolfgang Goethe-Universit\"at,\\ 60438 Frankfurt am Main, Germany\\
E-mail: \email{g.bergner@physik.uni-frankfurt.de}, \email{philipsen@th.physik.uni-frankfurt.de}}
\author{ J.~Langelage\\
Institute for Theoretical Physics, ETH Z\"urich, CH-8093 Z\"urich, Switzerland\\
E-mail: \email{ljens@phys.ethz.ch}}

\abstract{%
Effective Polyakov loop theories are a useful tool for an investigation of pure Yang-Mills theory and full QCD. A systematic derivation of the effective action can be done in a spatial strong coupling expansion. Quite accurate predictions for the deconfinement phase transition of Yang-Mills theory have been obtained in this approach. Besides the critical couplings, further observables can be measured in the effective theory. These provide additional tests for the reliability of the strong coupling approach and the truncation of the effective action. In this contribution we will present recent results for the free energy of the static quark-antiquark pair and the equation of state.
}

\FullConference{31st International Symposium on Lattice Field Theory - LATTICE 2013\\
		July 29 - August 3, 2013\\
		Mainz, Germany}

\begin{document}
\section{The effective Polyakov loop action approach with spatial strong coupling methods}
Some essential properties of finite temperature Yang-Mills theory can be represented in effective Polyakov line theories.
In recent years there have been several attempts for a derivation of these effective theories. 
%Especially for the confined phase and even up to the phase transition they allow for a simple and efficient description of the full gluodynamics.
Several non-perturbative approaches include a numerical derivation of the effective action \cite{Wozar:2007tz,Greensite:2013yd}. Apart from the numerical uncertainties  they also imply specific assumptions for the truncation of the effective action. 
Another disadvantage is that each value of the Yang-Mills coupling $\beta$  requires a new numerical computation of the coupling constants in the effective theory. In some cases even the form and truncation of the effective action is changed in this procedure.

Here we apply strong coupling methods for the derivation of the effective theory \cite{Langelage:2010yr}. This approach leads to a natural truncation and ordering of the operators in the effective action. We have computed up to a certain order in the expansion the analytic dependence of the effective couplings on $\beta$. Therefore no further assumptions or numerical input is needed in this approach.
On the other hand, it starts from the strong coupling region of the theory and is valid only below the confinement-deconfinement phase transition. Even below the transition the validity at larger $\beta$ has to be further investigated and confirmed for a given order
of the expansion.

One of the main prospects of effective descriptions is the inclusion of the fermionic degrees of freedom by means of a hopping parameter expansion. In principle the dynamics of QCD can then be described by a simple effective theory, which can be easily extended to finite chemical potential.
In the corresponding effective theory it is much simpler to handle the sign problem that is generic for the simulations at finite density \cite{Fromm:2011qi}.

In the current presentation we are considering the effective Polyakov line actions for pure Yang-Mills theory. We are investigating different observables for a detailed comparison of the effective 
theory derived in a strong coupling expansion with the full theory.  The effective Polyakov line action is obtained when the spatial link variables $U_i$ are integrated out in the path integral of 
Yang-Mills theory on the lattice,
\begin{equation}
e^{-S_{\text{eff}}[L]}\equiv \int [d U_i]\; 
\prod_p e^{\frac{\beta}{6} \Tr \left(U_p+U_p^\dag \right)}\; , \quad  Z=\int [dL] e^{-S_{\text{eff}}[L]}\; .
\end{equation}
In this equation the Wilson action with the product of all plaquette contributions $U_p$ is used and the gauge group is SU(3).
In the end the result depends only on gauge invariant products of the temporal links, the Polyakov loops, $L({\bf x})=\Tr\prod_\tau U_0({\bf x})$.
In principle, these Polyakov loops appear in all irreducible representations of the gauge group in the effective action.
However, the influence of higher representations is suppressed in our strong coupling approach.
The theory is reduced from a four dimensional gauge theory to a simple three dimensional theory.
This theory can be easily simulated with Monte-Carlo methods and precise results for different observables can be obtained in this way.
The simulation effectively corresponds to the computation of the remaining part of the path integral.

As stated above there are  numerical strategies for the derivation of the effective Polyakov loop theories \cite{Wozar:2007tz,Greensite:2013yd}.
Once the effective action is used as an input for numerical simulations, these strategies correspond just to a splitting of the integration of spatial and temporal link variables. 
Compared to the full path integral they are only advantageous if assumptions can be made concerning the dependence of the effective theory on the parameters of the full theory. If, for example, from the effective theory at zero chemical potential the form at nonvanishing chemical potential can be deduced, as argued in \cite{Greensite:2012xv}, this approach is beneficial.

Here we integrate out spatial links in a strong coupling expansion. In contrast to the straight-forward implementation of this idea, as done in \cite{Polonyi:1982wz}, it includes the resummation of certain terms in the expansion.
To achieve a better convergence, we use the coefficient of the character of the fundamental representation $u(\beta)=u/18+O(\beta^2)$ as a new expansion parameter, instead of $\beta$.
The leading term of the effective action is obtained from the contribution of two Polyakov lines filled with plaquettes in the fundamental representation.
\begin{figure}
\begin{center}
\includegraphics[height=5cm]{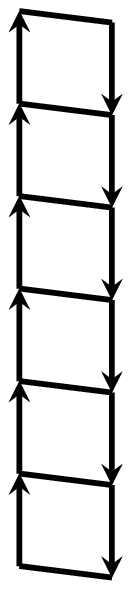}
\raisebox{2cm}{$+$}
\includegraphics[height=5cm]{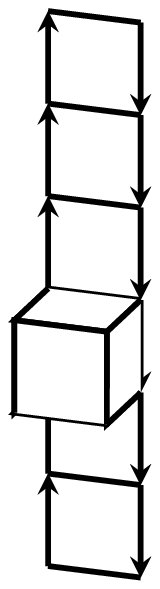}
\raisebox{2cm}{$+$}
\includegraphics[height=5cm]{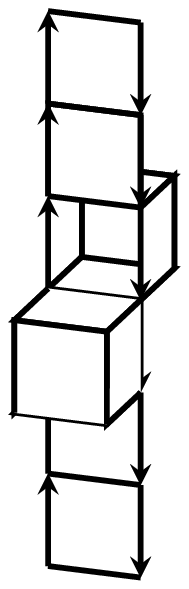}
\raisebox{2cm}{$\longrightarrow$}
\includegraphics[height=5cm]{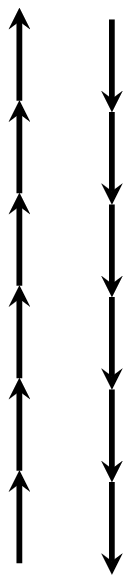}
\hspace*{2cm}
\includegraphics[height=5cm]{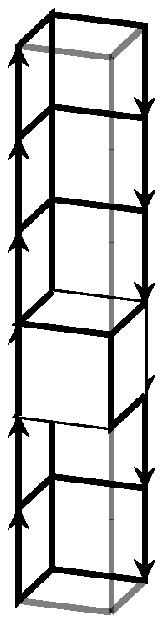}
\raisebox{2cm}{$\longrightarrow$}
\includegraphics[height=5cm]{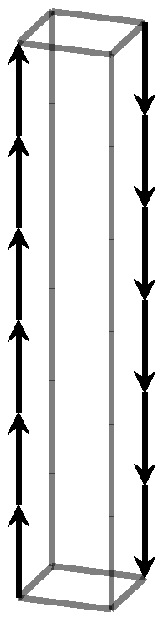} 
\end{center}
\caption{A graphical representation of the contributions in the strong coupling expansion and the corresponding terms in the effective action.
The first term is the interactions of two nearest neighbour Polyakov lines and the second one corresponds to the interaction of next-to nearest neighbours with distance $\sqrt{2}$ on the lattice. }
\label{fig:effg}
\end{figure}
Figure \ref{fig:effg} illustrates how higher order contributions arise in terms of additional contributions to the coupling constants of the nearest neighbour interaction and next to nearest neighbour interactions.

The strong coupling derivation of the effective action orders the interactions according to their leading power in the expansion.
The long range interactions and higher representations come with higher orders of $u$   ($u^{2N_t+6}$; $u^{2N_t+2}$) and are suppressed,
\begin{equation}
 S_{\text{eff}}=\lambda_1 S_{\text{nearest neighbors}}+\lambda_2 S_{\text{next to nearest
neighbors}}+\ldots\; .
\end{equation}
In this contributions we consider only the first term, the nearest neighbour interaction. 
It receives corrections at higher orders in $u$ that can be resummed to
\begin{equation}
\lambda_1(u)=u^{N_t} \exp(N_t P_{N_t}(u)) \; ,
\label{eq:lofu}
\end{equation}
where $P_{N_t}(u)$ is a polynomial to order $u^{10}$ \cite{Langelage:2010yr}.
Furthermore, the higher powers of this interaction can be summed in a logarithmic form of the effective action
\begin{equation}
S_{\text{nearest neighbors}}=\sum_{<ij>} (2\lambda_1 \Re L_i L_j^* -\frac{1}{2}(2\lambda_1 \Re L_i L_j^*)^2+
\ldots) =\sum_{<ij>} \log(1+2\lambda_1 \Re L_i L_j^*) \; .
\end{equation}

A numerical simulation of the effective theory includes besides the strong coupling contributions the nonperturbative effects of the 
Polyakov line dynamics. As in SU(3) Yang-Mills theory a first order phase transition is observed, corresponding to a spontaneously broken 
centre symmetry. In the investigated range of $N_t\leq 16$ there is only a few percent difference between the critical $\beta$ value of the transition in the effective theory and $\beta_c$ in full Yang-Mills theory. A convergence of the critical temperature of the effective theory towards the full theory has been observed in the continuum limit \cite{Fromm:2011qi}. At a finite lattice spacing the difference of the two critical values of $\beta_c$, and correspondingly $T_c$, must be taken into account.

\section{Comparison of Polyakov line correlators}
The Polyakov line correlators, $\langle L(\vec{0}) L^\dag (\vec{R}) \rangle$, are simple observables that can be used for a comparison of the effective theory and the full theory. Efficient algorithms \cite{Luscher:2001up} offer a precise measurement of them in the Yang-Mills theories.
Moreover they encode interesting physical properties of the system. The unrenormalized free energy of a static quark-antiquark pair is obtained from
\begin{equation}
\langle L(\vec{0}) L^\dag (\vec{R}) \rangle=\exp(-F(|\vec{R}|,T)/T)\; .
\label{eq:fofc}
\end{equation}
In the continuum limit this result depends only on the distance $|\vec{R}|$ and $T$ and is invariant under rotations.
This rotational invariance is broken at a finite lattice spacing. The restoration of the symmetry in the continuum limit is an important property of Yang-Mills theory.
 
\begin{figure}
 \begin{center}
\begin{minipage}{7cm}
\begin{center}
{\tiny $\beta=5.0$} \\
 \includegraphics[width=7cm]{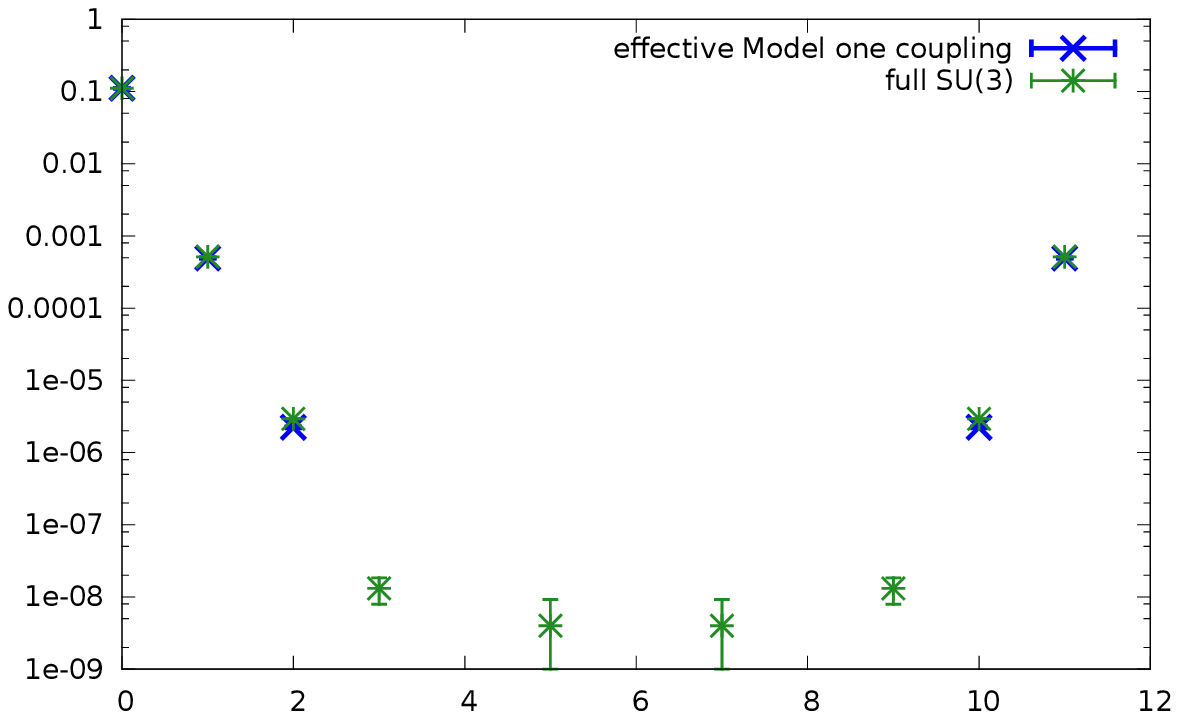}
\end{center}
\end{minipage}
\begin{minipage}{7cm}
\begin{center}
{\tiny $\beta=5.4$}\\
 \includegraphics[width=7cm]{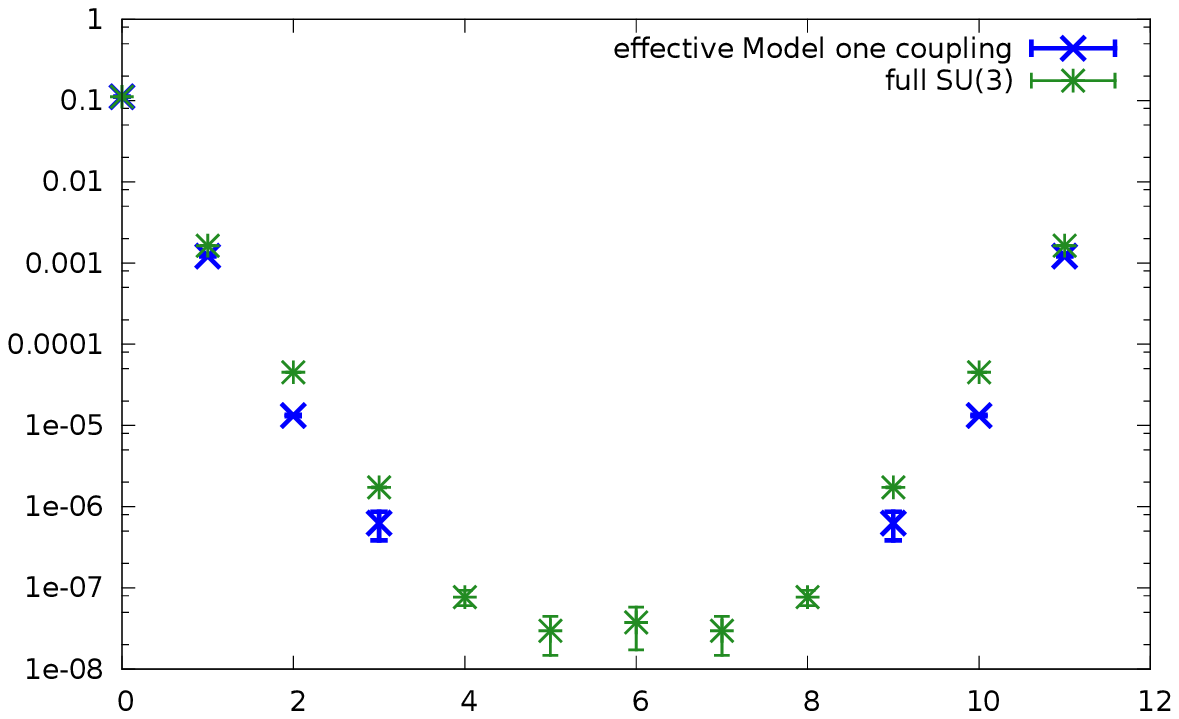}
\end{center}
\end{minipage}
\end{center}
\caption{The Polyakov line correlators of the effective theory compared with those of the full Yang-Mills theory. 
The comparison is done at two different values of $\beta$. Due to the large statistical errors some of the points for the correlators of the effective theory have been excluded from the plot. Only on-axis distances are included. The lattice size is $6\times 12^3$, corresponding to $12^3$ in the effective theory.}
\label{fig:pcor}
\end{figure}

A comparison of the correlator in full Yang-Mills theory and in the effective theory is shown in Figure \ref{fig:pcor}. Only on-axis distances are included in this figure. Due to the exponential error reduction for this observable in Yang-Mills theory the statistical error is lower than in the effective theory, despite the larger statistic.
At lager values of $\beta$ the difference between the effective theory result and the full theory increases, but there is still a reasonable agreement. 

\begin{figure}
\begin{center}
\begin{minipage}{7cm}
\begin{center}
{\tiny $\beta=5.0$} \\
 \includegraphics[width=7cm]{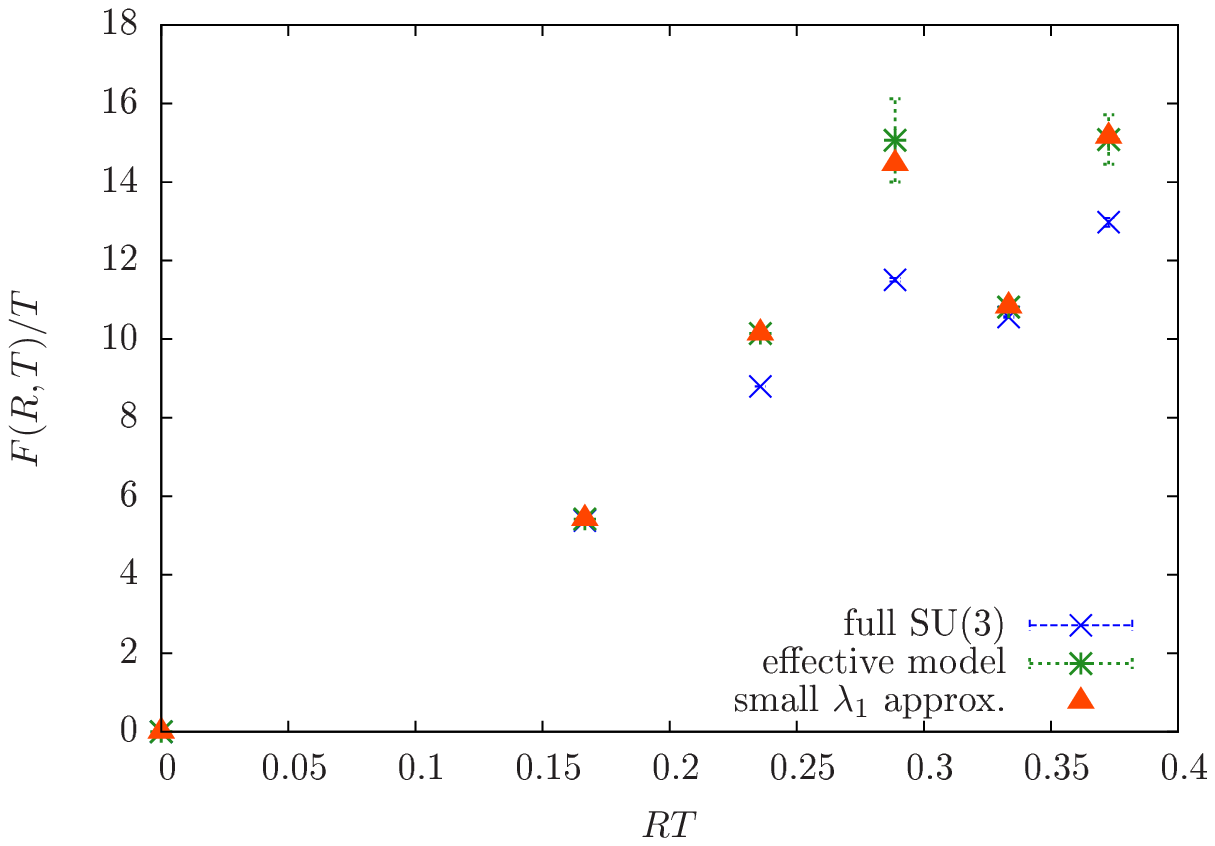}
\end{center}
\end{minipage}
\begin{minipage}{7cm}
\begin{center}
{\tiny $\beta=5.5$}\\
 \includegraphics[width=7cm]{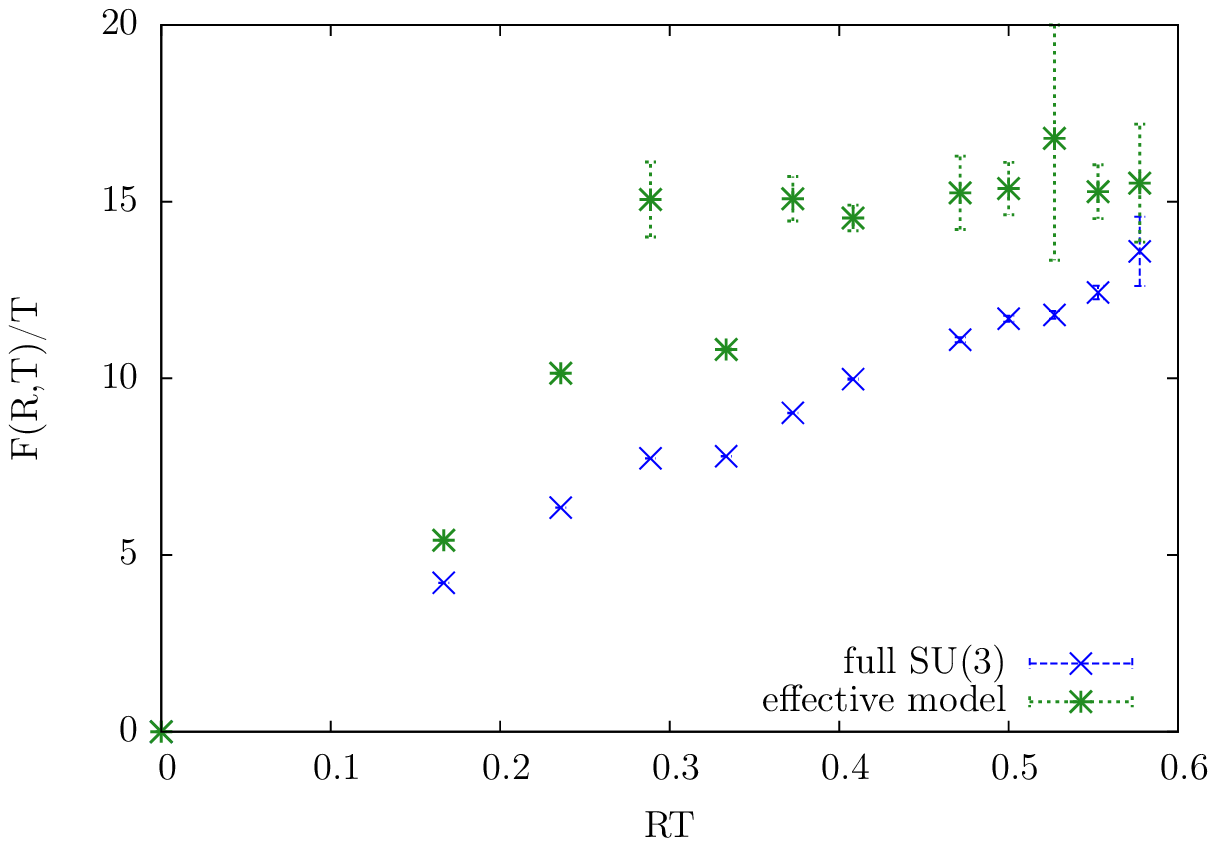}
\end{center}
\end{minipage} 
\end{center} 
\caption{The free energy of a static quark-antiquark pair determined form the correlators according to Equation \protect\ref{eq:fofc}.
The results of the effective theory and full Yang-Mills theory is shown for two different values of $\beta$. At the smaller $\beta$ value the comparison includes also the analytic results of an expansion in $\lambda_1$. The lattice size is the same as in Figure \protect\ref{fig:pcor}.}
\label{fig:vgloff}
\end{figure}
A more challenging analysis can be done when also the off-axis correlations are included in the comparison. 
As shown in Figure \ref{fig:vgloff} the expected breaking of the rotational invariance at small $\beta$ is 
clearly visible in the off-axis correlator. This observation is confirmed by the analytic result for the correlator to the lowest orders in $\lambda_1$,
\begin{equation}
 F(R,T)/T=\log(9)-\log\left[\lambda_1^{\text{dist}(R/a)}(N_1(R/a)+N_2(R/a)\lambda_1^2+O(\lambda_1^4))\right]\; ,
\end{equation}
where $\text{dist}(R/a)$ is the the shortest distance along the lattice links and $N_i$ is a combinatorial factor.

At a given value of $\beta$ the effective theory shows a more pronounced breaking of the rotational symmetry than the full Yang-Mills theory.
This effect becomes clearly visible in the vicinity of the phase transition, where a larger deviation between the off-axis correlators in the full theory and the effective theory can be observed.

The rotational invariance is restored at larger values of $\lambda_1$ in the effective theory.
It happens below the critical coupling $(\lambda_1)_c$ but above the value $\lambda_1(u(\beta_c))$, where $\beta_c$ is the critical $\beta$ in the full theory and the map of the couplings in Equation \ref{eq:lofu} has been applied.
Inverting this map one obtains $\beta((\lambda_1)_c)$ and, correspondingly,  a temperature $(T_c)_{\text{eff}}(\beta((\lambda_1)_c))$ corresponding to the phase transition in the effective theory\footnote{The lattice spacing $a$ is determined by the scale setting of the full Yang-Mills theory $a(\beta)$. The formula given in \cite{Necco:2001xg} is used for this scale setting.}.

\begin{figure}
 \begin{center}
  \begin{minipage}{7cm}
\begin{center}
{\tiny full Yang-Mills theory, $T=98\% T_c$} \\
 \includegraphics[width=7cm]{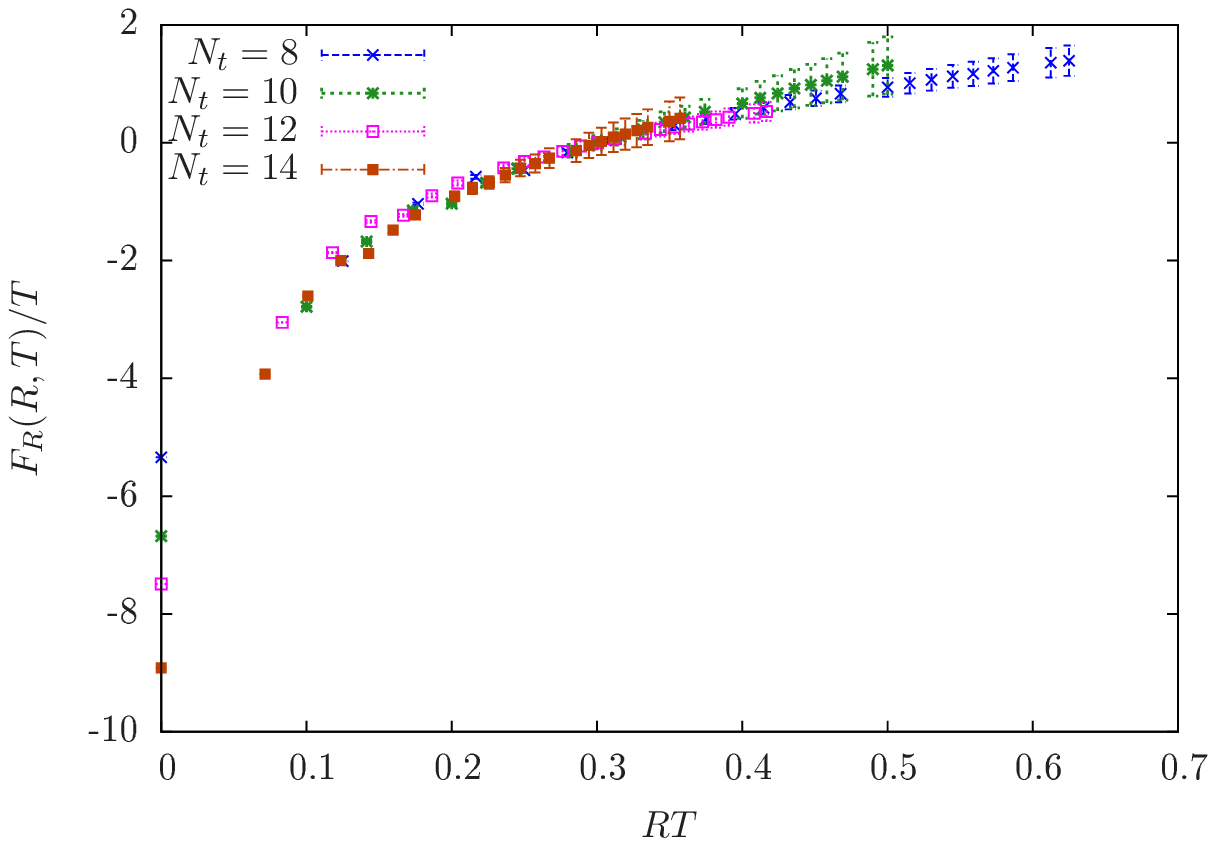}
\end{center}
\end{minipage}
\begin{minipage}{7cm}
\begin{center}
{\tiny effective Polyakov loop theory, $T=98\% T((\lambda_1)_c)$} \\
 \includegraphics[width=7cm]{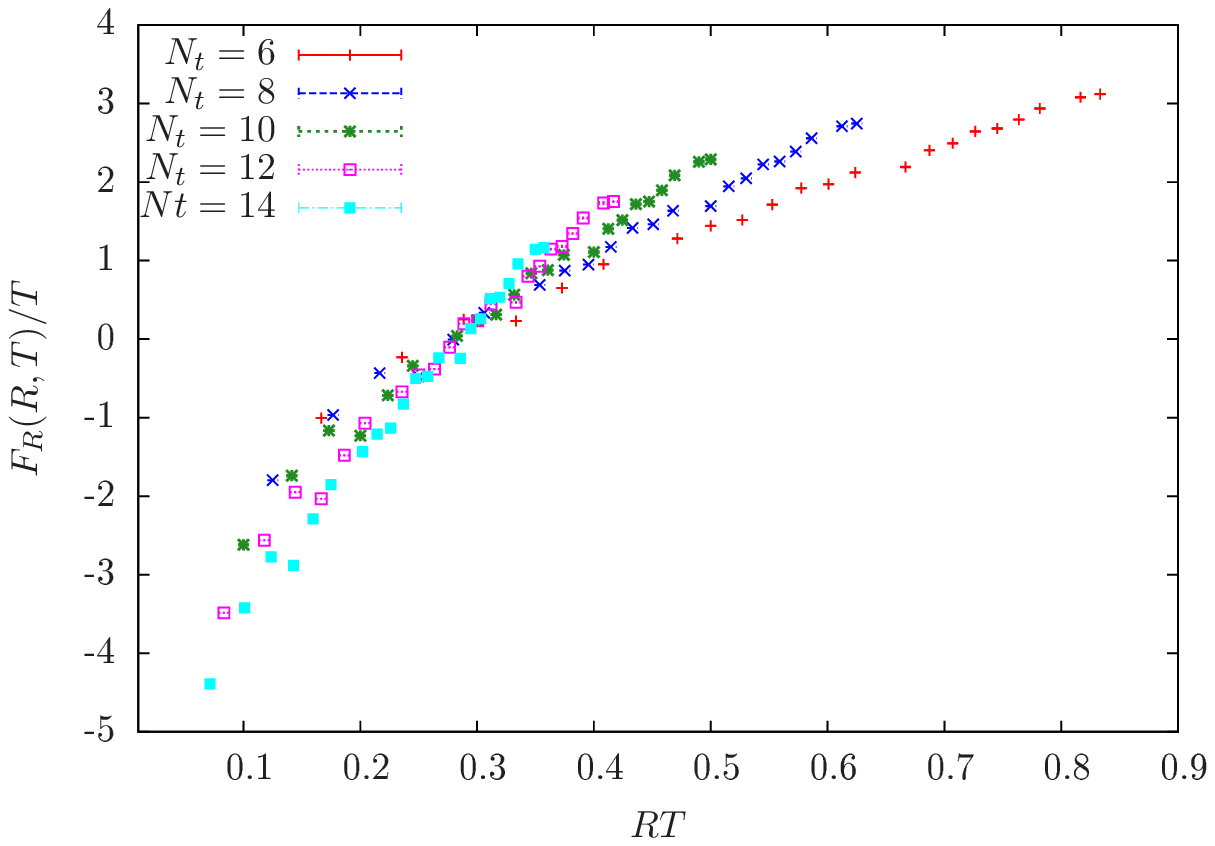}
\end{center}
\end{minipage}
 \end{center}
\caption{The renormalized free energy as a function of the physical distance $R$ in units of the temperature.
The ratio of temperature over critical temperature is fixed to $98\%$ in both theories with the corresponding critical temperature.
The simulations were done on a $N_t\times (2N_t)^3$ lattice for various $N_t$ in the Yang-Mills case. In the effective theory the lattice size is fixed to $32^3$.}
\label{fig:TTC}
\end{figure}
The different $\beta$ values for the restoration of the rotational symmetry suggests a comparison of the correlators in the vicinity of the respective phase transitions.
In addition the bare value of $F/T$ still contains a zero temperature part that diverges in the continuum limit.
For a consistent definition of the renormalized free energy, $F_R/T$, this part needs to be subtracted.
We have done this subtraction by setting  $F_R(R_0,T)/T$ to zero for a given physical distance $R_0T$.

Figure \ref{fig:TTC} summarizes the results of this comparison.  The data of the full Yang-Mills theory shows only small deviations from the universal scaling towards the continuum limit. The values for different lattice spacings are nearly on top of each other. 
The correlators of the effective theory in this close vicinity of the phase transition $T((\lambda_1)_c)$ show a restoration of the rotational symmetry. At larger distances $F_R/T$ depends almost linearly on $RT$. All these general features are in good agreement with full Yang-Mills theory. In a quantitative comparison there are, however, some considerable differences between the effective theory and the full theory.
The slope of the correlator at larger distances, corresponding to the string tension, is larger in the effective theory.
Furthermore a comparison of different lattice spacings indicates that the string tension grows towards the continuum limit.
The accurate representation of the correlators and its exponential decay in  the effective theory seems to require interactions terms at larger distances.
%%%%%%%%%%%%%%%%%%%%%%%%%%%%%%%%%
\section{Thermodynamics of the SU(3) effective Polyakov line theory}
 \begin{figure}
  \begin{center}
 \includegraphics[width=8cm]{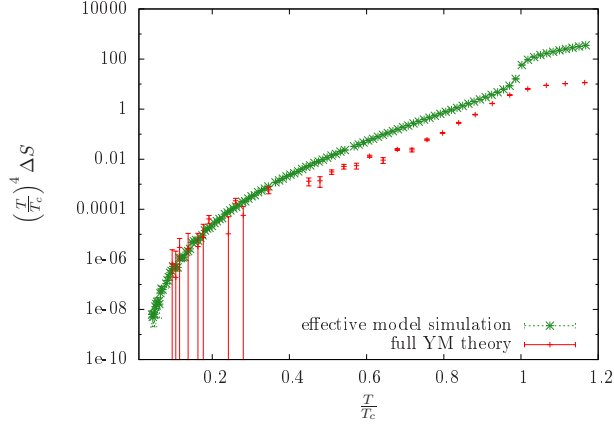}   
  \end{center}
\caption{The primary thermodynamic quantity $\Delta S$ as a function of $T/T_c$ in full Yang-Mills theory and the effective theory.}
\label{fig:deltas}
 \end{figure}
The investigations of the equation of state on the lattice start from the interaction measure $\Delta S$ (see e.~g.\ \cite{Boyd:1996bx}),
 \begin{equation}
  \Delta S\equiv -\frac{d}{d \beta}\frac{f}{T^4}=\frac{1}{V T^3}\frac{d}{d \beta}\log Z\; .
 \end{equation}
It is derived from differences of simple plaquette expectation values. The thermodynamic quantities, e.~g.\ the pressure, are obtained from this
primary observable.
For a comparison we measure the same quantity  in the effective theory. 
The derivation is based on the analytic dependence of $\lambda_1$ on $\beta$.
Similar to the full theory, $\Delta S$ is given by differences of simple expectation values.
A comparison between the results from the effective theory and the full theory is shown in Figure \ref{fig:deltas}.
 It indicates that the two results converge in the small $\beta$, i.~e.\ strong coupling region.
%%%%%%%%%%%%%%%%%%%%%%%%%%%%%%%%%
\section{Conclusions}
We have summarized non-perturbative tests of the effective Polyakov loop action for Yang-Mills theory derived by means of a spatial strong coupling expansion. In particular we used its simplest form with only next neighbour interactions.

When comparing correlation functions, then at large distances and towards the continuum limit higher orders of the expansion and, in particular, interactions over larger distances become important.
The approach is hence limited to a certain range of lattice spacings. 
The most important concern is whether an intermediate scaling region can be identified 
and continuum results can already be extrapolated from the effective theory.

On the other hand, the critical coupling for the phase transition as well as bulk thermodynamic properties are well described by the one-coupling effective theory.
%%%%%%%%%%%%%%%%%%%%%%%%%%%%%%%%%%%%%%%%%%%%%%%%%%%%%%%%%%%%%%%
\section*{Acknowledgments}
GB and OP are acknowledge support by the German BMBF, No. 06FY7100.
JL is supported by SNF grant 200020\_137920.
%%%%%%%%%%%%%%%%%%%%%%%%%%%%%%%%%%%

\end{document}